\documentclass{aastex62}
\usepackage{graphicx}
\usepackage{array,multirow}
\usepackage{mathtools}

\received{--}
\revised{-- }
\accepted{ --}
\submitjournal{ApJ}

\shorttitle{Variation of non-thermal line widths}
\shortauthors{Pant et al.}

\begin{document}

\title{\bf{\large{Revisiting the relation between nonthermal line widths and transverse MHD wave amplitudes}}}

\correspondingauthor{Vaibhav Pant}
\email{vaibhav.pant@kuleuven.be, vaibhavpant55@gmail.com}

\author{Vaibhav Pant}
\affiliation{Centre for mathematical Plasma Astrophysics, Department of Mathematics, KU Leuven, Celestijnenlaan 200B, Leuven, Belgium}

\author{Tom Van Doorsselaere}
\affiliation{Centre for mathematical Plasma Astrophysics, Department of Mathematics, KU Leuven, Celestijnenlaan 200B, Leuven, Belgium}

\begin{abstract}
Observations and 3D MHD simulations of the transverse MHD waves in the solar corona have established that true wave energies hide in the nonthermal line widths of the optically thin emission lines. 
This displays the need for a relation between the nonthermal line widths and transverse wave amplitudes for estimating the true wave energies. In the past decade, several studies have assumed that the root mean square (rms) wave amplitudes are larger than nonthermal line widths by a factor of $\sqrt{2}$. However, a few studies have ignored this factor while estimating rms wave amplitudes. Thus there appears to exist a discrepancy in this relation.  
In this study, we investigate the dependence of nonthermal line widths on wave amplitudes by constructing a simple mathematical model followed by 3D MHD simulations. We derive this relation for the linearly polarised, circularly polarised oscillations, and oscillations excited by multiple velocity drivers. We note a fairly good match between mathematical models and numerical simulations. We conclude that the rms wave amplitudes are never greater than the nonthermal line widths which raises questions about earlier studies claiming transverse waves carry enough energy to heat the solar corona.
\end{abstract}


\section{Introduction}
The mechanism of heating of the solar corona can be broadly divided into two categories. Heating due to the dissipation of waves and heating by current dissipation due to the magnetic reconnection \citep{2003A&ARv..12....1W, 2012RSPTA.370.3217P}. Propagation of MHD waves and their contribution to the coronal heating has been investigated for many decades \citep{2007SoPh..246....3B,2006SoPh..234...41K,2014ApJ...795..111H,2015RSPTA.37340261A}. Perhaps, the earliest signatures of Alfv\'en(ic) waves in the solar atmosphere are the nonthermal broadening of the optically thin emission lines due to the unresolved wave amplitudes \citep{1973ApJ...181..547H,1976ApJ...205L.177D, 1976ApJS...31..417D}.
After the launch of {\it Solar and Heliospheric Observatory} (SOHO) and {\it Hinode}, the non-thermal broadening of the spectral lines has been unambiguously observed and reported in the solar atmosphere \citep{1976ApJ...205L.177D, 1976ApJS...31..417D, 1976ApJS...31..445F, 1990ApJ...348L..77H, b98,1998SoPh..181...91D,2005A&A...436L..35O,b2009,2012ApJ...753...36H}. The non-thermal broadening is observed to vary with heights above the solar atmosphere \citep{1976ApJ...205L.177D,1998SoPh..181...91D,2012ApJ...753...36H}. However, the nature of the variation is different in different regions of the Sun \citep{2019A&A...631A.163D,2019A&A...627A..62G}. These studies assume that the nonthermal broadening is produced by the unresolved Doppler velocity amplitudes in time due to the presence of MHD waves in the solar corona. To complicate matters, several physical processes other than MHD waves affect nonthermal line widths in the solar atmosphere. 
Plasma upflows along the coronal loops and plumes can cause nonthermal broadenings at the footpoints of these structures \citep{2010ApJ...722.1013D,2011ApJ...727L..37T,2012ApJ...759..144T}. Large scale upflows (possibly responsible for the solar wind) in open magnetic field regions such as coronal holes can also contribute towards broadening of a spectral line \citep{2011ApJ...727....7M,2011ApJ...736..130T}. 

Both numerical simulations and observations have suggested the presence of counter-propagating waves in the solar atmosphere \citep{2009ApJ...697.1384T,2015NatCo...6E7813M,2017ApJ...849...46V}. Such counter-propagating waves may lead to turbulence (also termed as Alfv\'en wave turbulence; AWT) and non-linear cascade of energy to small spatial scales. This causes nonthermal broadening of emission lines \citep[see Figure 8 in][]{2017ApJ...849...46V}. Apart from AWT, a new mechanism for generating turbulence by Alfv\'enic waves propagating in transversely inhomogeneous plasma is reported by \citet{2017NatSR...714820M}. This type of turbulence is termed as uniturbulence because it relaxes the need of counter-propagating waves to generate turbulence. A unidirectionally propagating wave in the presence of transverse inhomogeneities cause self-deformation and non-linear cascade of energy \citep{2019ApJ...882...50M}. Since the solar corona is highly structured (transversely inhomogeneous), uniturbulence is inevitable and could play an important role in the nonthermal broadening especially in the open magnetic field regions where waves are predominantly unidirectional. 

Finally, the superposition of structures swaying in different directions along the line-of-sight (LOS) also broaden an optically thin emission line  \citep{2012ApJ...761..138M,2019ApJ...881...95P}. Recently, \citet{2019ApJ...881...95P} studied the role of LOS superposition and  uniturbulence in explaining the observed spectral properties of transverse MHD waves propagating in the coronal holes. These authors have performed ideal 3D MHD simulations and reproduced the observed large nonthermal line widths, small resolved Doppler shifts, and a wedge-shaped correlation between them. These authors also reported that true wave amplitudes (and thus energy) are hidden in the nonthermal line widths.

In the present study, we ignore the effects of flows, (uni)turbulence and assume that the nonthermal broadening is caused by unresolved Doppler shifts generated due to the transverse MHD waves in the solar atmosphere. Often, the energy content of a transverse wave is computed by estimating the root mean square (rms) velocity of the wave amplitude \citep{1981SoPh...70...25H}. Further, rms wave amplitudes can be estimated by measuring nonthermal broadening of optically thin emission lines. Thus, the observed magnitude of the nonthermal line widths can provide an estimate of the energy carried by transverse waves. Therefore, it is imperative to understand the relation between nonthermal line widths and rms wave amplitudes. In fact, such a relation was used to compute the Alfv\'en(ic) wave energy flux \citep{1990ApJ...348L..77H,b98,1998SoPh..181...91D}. These studies assumed that the rms wave amplitude, $v_{rms}$, is related to the nonthermal line width, $\sigma_{nt}$ as $v_{rms}=\alpha~\sigma_{nt}$, considering different polarisations and direction of propagation of transverse waves relative to a LOS \citep{1990ApJ...348L..77H}. Here, $\alpha$ was assumed to be approximately $\sqrt{2}$. A similar relation was later reported by \citet{1998SoPh..181...91D} accounting for two degrees of freedom for an Alfv\'en wave. Since then this relation has been used extensively in many studies for estimating the energy carried by the Alfv\'en(ic) wave using observed values of the nonthermal line widths in the solar corona \citep{2005A&A...436L..35O,b2009,2012ApJ...753...36H}.\\
Surprisingly, a few studies have ignored $\alpha$ while estimating the wave energies \citep{1998ApJ...503..475T,1998ApJ...505..957C,2012ApJ...751..110B,2012ApJ...759..144T}. \citet{1998ApJ...505..957C} argued that $v_{rms}=\sigma_{nt}$ in a pure Alfv\'en wave because only the directions perpendicular to the wave motion contribute to the energy transport. On other hand, \citet{1998ApJ...503..475T} argued that the degrees of freedom in the Alfv\'enic waves or turbulence is 2, therefore $v_{rms}=\sigma_{nt}$. Till today, there is no convincing explanation which one of these relations should be used to estimate the wave amplitude and hence energy. This serves as a motivation for the study described in this paper. We present the relations between $v_{rms}$ and $\sigma_{nt}$ for different velocity drivers by constructing a simple mathematical model. We confirm the results of the mathematical model with MHD simulations. Finally we discuss the implication of our study for studying the role of waves on coronal heating.

\section{Mathematical model}
\label{sec2}
\begin{figure}[!ht]
\centering
\includegraphics[scale=0.6, angle=90]{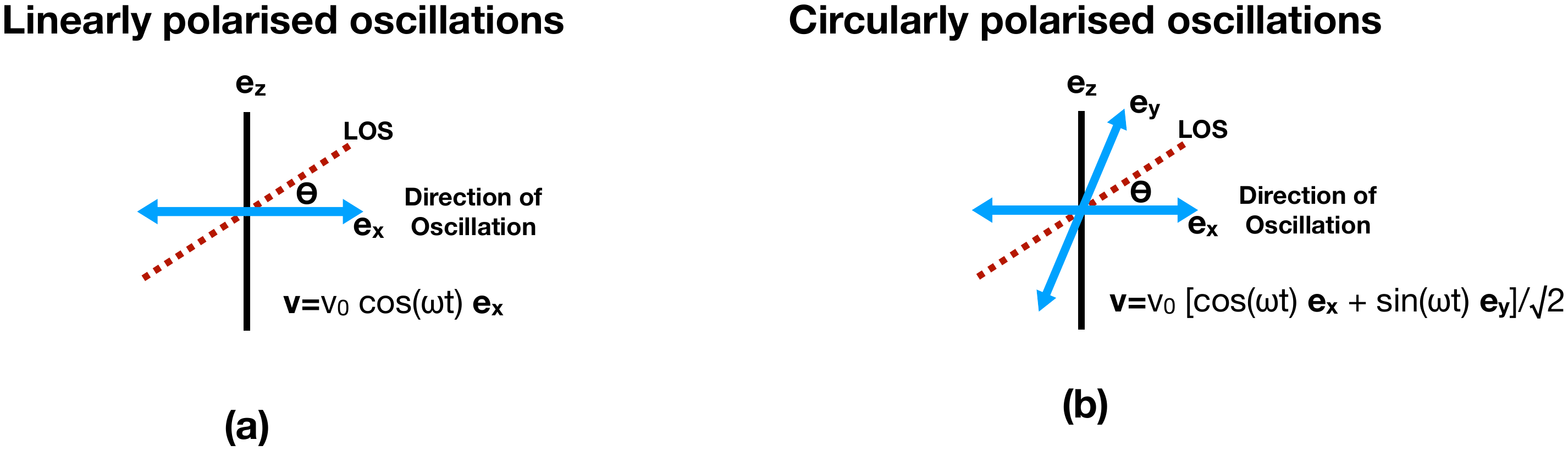}
\caption{Schematic diagrams for (a) linearly polarised and (b) circularly polarised oscillations. The dashed curves represent the direction of the line-of-sight (LOS). $\theta$ represents the angle between LOS and unit vector in $x$-direction. For circularly polarised oscillations, velocity of oscillations have phase difference of $\pi$/2 along $x$ and $y$ directions.}
\label{fig1}
\end{figure}

We assume a uniform plasma at temperature $T$ oscillating along the LOS with a period $P$ and velocity $v$, as shown schematically in Figure~\ref{fig1} (a). Further assuming that the emitting plasma is in thermal equilibrium, the  shape of a spectral line ($G(\lambda)$) at an instant $t$ is given by the following relation \citep{10.3389/fspas.2016.00004}.  
\begin{equation}
    G(\lambda) = \frac{1}{ \sigma_{w} \sqrt{2 \pi}} \exp\left(-\frac{1}{2\sigma_{w}^2} \left(\lambda - \lambda_{0}(1 \pm \frac{v}{c})\right)^2\right).
\label{eq1a}
\end{equation}
Here $v$ is the velocity of the emitting plasma along LOS, $\lambda_{0}$ is the central wavelength of the emission, and $\lambda_{0}(1 \pm \frac{v}{c})$ is the wavelength shift due to the velocity of emitting plasma along the LOS.  We assume the Doppler broadening (due to the temperature of the plasma) to be the dominant broadening mechanism when plasma is at rest. The width of the Gausssian shaped spectral line, $\sigma_{w}$, is defined as $\sqrt{2k_{b}T/M} \lambda_{0}/c \sqrt{2}$. Note, $k_{b}$ is the Boltzmann constant, $M$ is the mass of the emitting ion in the plasma, and $c$ is the speed of light. In some studies $\sqrt{2k_{b}T/M}$ is termed as thermal line width or exponential line width ($\sigma_{1/e}$). For simplicity, we choose $\sigma=\sqrt{2k_{b}T/M}/\sqrt{2}$ such that $\sigma_{w}=\sigma \lambda_{0}/c$. Inserting the expression for $\sigma_{w}$ in Equation~\ref{eq1a}, we get
\begin{equation}
    G(\lambda)=\frac{c}{ \sigma \lambda_{0} \sqrt{2 \pi}} \exp\left(-\frac{1}{2\sigma^2} \left(\frac{(\lambda - \lambda_{0})c}{\lambda_{0}} \pm v\right)^2\right).
    \label{eq1b}
\end{equation}
Often in literature, for simplicity, we assume the dependence of $G(\lambda)$ on velocity instead of wavelength. To convert wavelength to velocity, we define a new variable $\mathfrak{v}$ such that when $\lambda$ is replaced by $\lambda_{0}(1+ \frac{\mathfrak{v}}{c} )=\lambda'$, equation~\ref{eq1b} reduces to
\begin{equation}
    G(\lambda ')=\frac{c}{ \sigma \lambda_{0} \sqrt{2 \pi}} \exp\left(-\frac{1}{2\sigma^2} \left(\mathfrak{v} \pm v\right)^2\right).
    \label{eq1c}
\end{equation}
It should be noted that $\frac{c}{ \sigma \lambda_{0} \sqrt{2 \pi}}$ is a normalisation constant such that $\int G(\lambda')d\lambda'=1$. This constant will have no effect on the analysis presented in this paper. Equation~\ref{eq1c} can be reformulated in terms of variable $\mathfrak{v}$ as follows,
\begin{equation}
    G(\mathfrak{v})=\frac{1}{ \sigma \sqrt{2 \pi}} \exp\left(-\frac{1}{2\sigma^2} \left(\mathfrak{v} \pm v\right)^2\right).
    \label{eq1d}
\end{equation}
such that $\int G(\mathfrak{v})d\mathfrak{v}=1$. Equation~\ref{eq1d} gives the shape of a spectrum, emitted by the plasma moving with velocity $v$ along the LOS, in the Doppler shifted velocity coordinate ($\mathfrak{v}$) such that if the emitting ion is at rest, the Gaussian spectrum will be centered around zero. Henceforth, we will employ Equation~\ref{eq1d} instead of Equation~\ref{eq1a} for further analysis.

\subsection{Oscillations along the LOS and effect of finite exposure time}
\label{sec2.1}
We assume a harmonic plane wave propagating along the background magnetic field. Following the analysis presented in \citet{walker}, the velocity of a transverse wave can be assumed to be $\propto$ $e^{-i\omega t}$. In general, for a linearly polarised oscillation, ${\bf v}=\frac{1}{2}[v_{0}e^{i\omega t} + v^{*}_{0}e^{-i\omega t}]{\bf e_{x}}$. Taking only real amplitudes of the wave velocity, we get ${\bf v}=v_{0}\cos{\omega t}~{\bf e_{x}}$. Here ${\bf e_{x}}$ is the unit vector along the $x$ axis.\\
Now, let us assume that the angle between LOS and $e_{x}$ as shown in Figure~\ref{fig1} vanishes. Thus, the LOS is in the ${\bf e_{x}}$ direction. Further we assume that wave amplitudes are unresolved in time either due to the high frequency nature of the oscillations or due to a large exposure time, $t$, of an instrument such as SUMER ($\sim$300 s). In such a scenario, the spectrum recorded by the spectrograph will be approximately given by the following relations.   
\begin{equation}
\begin{aligned}
    \big \langle G(\mathfrak{v},t') \big \rangle _{t}=\int_{0}^{t} \frac{1}{t \sigma \sqrt{2 \pi}} \exp(-\frac{(\mathfrak{v}-v_{0}~\cos(\omega t'))^{2}}{2\sigma^{2}}) dt',\\
    =\Big(\int_{0}^{fP} \frac{1}{t \sigma \sqrt{2 \pi}} \exp(-\frac{(\mathfrak{v}-v_{0}~\cos(\omega t'))^{2}}{2\sigma^{2}}) + \int_{fP}^{t} \frac{1}{t \sigma \sqrt{2 \pi}} \exp(-\frac{(\mathfrak{v}-v_{0}~\cos(\omega t'))^{2}}{2\sigma^{2}}) \Big) dt'.
\end{aligned}
\label{eq2}
\end{equation}
Here, $P$ is the period of the oscillations and $f$ is a positive integer such that $fP \le t < (f+1)P$. If $f \gg 1$,  we get $t~\sim~fP$. Thus, the contribution of the second term in equation~\ref{eq2} towards the estimation of $\big \langle G(\mathfrak{v},t') \big \rangle _{t}$ will be much less compared to the first term. Therefore, we can neglect the second term in equation~\ref{eq2}. This means when the exposure time of an instrument is much larger than the period of oscillations, averaging the spectrum over one period is a good assumption. Note that we normalise equation~\ref{eq2} by the exposure time, t. The normalisation will not affect the estimation of the nonthermal line widths. We solve equation~\ref{eq2} analytically, assuming $v_{0}\ll \sigma$ (limiting case). We find that the nonthermal line width, $\sigma_{nt}$, of the period averaged spectrum ($\big \langle G(\mathfrak{v},t') \big \rangle _{t}$) is approximately equal to the wave amplitude $v_{0}$ (see appendix~\ref{appa} for the derivation). Further, we compare $\sigma_{nt}$ with the rms wave velocity computed using the following relation:

\begin{equation}
v_{rms}=\sqrt{\frac{1}{P} \int_{0}^{P} {\bf v\cdot v^{*}}dt}.
\label{eq3}
\end{equation}
For linearly polarised oscillations, ${\bf v\cdot v^{*}}=v_{0}^2~\cos^2(\omega t)$. Thus we find that, $\sigma_{nt}\sim\sqrt{2}~v_{rms}$.
Note that, if $v_{0}$ is comparable to $\sigma$, we can no longer ignore higher order terms of $v_{0}$ (see appendix~\ref{appa}). Thus, we need to solve equation~\ref{eq2} using numerical integration.\\
Next, we solve equation~\ref{eq2} using realistic wave amplitudes observed in the solar corona. Since the mean value of the transverse wave amplitude in the coronal holes is 11$\sqrt{2}$~km~s$^{-1}$ \citep{2015NatCo...6E7813M}, we numerically integrate equation~\ref{eq2} assuming $v_{o}$~=11$\sqrt{2}$~km~$s^{-1}$ and $\sigma$=19/$\sqrt{2}$~km~$s^{-1}$ such that the thermal width ($\sigma_{1/e}$) is 19~km~$s^{-1}$ (for 1.2 MK plasma). Throughout the manuscript, we use these values to perform the numerical integration. It is worth mentioning here that $v_{o}$ is comparable to the thermal line width, $\sigma_{1/e}$ but larger than $\sigma$. The period averaged spectrum, $\big \langle G(\mathfrak{v},t') \big \rangle _{t}$, obtained by numerical integration is shown in Figure~\ref{fig2} (a) in blue. Next, we fit a Gaussian curve to $\big \langle G(\mathfrak{v},t') \big \rangle _{t}$ and estimated the nonthermal line width, $\sigma_{nt}=\sqrt{\sigma_{1/e} '^{2} - \sigma_{1/e}^{2}}$. Here, $\sigma_{1/e}'$ is the exponential line width of the best-fit Gaussian curve shown in red.\\
In this case, we find that $\sigma_{nt}/v_{0}\sim$1.1 or $\sigma_{nt}/v_{rms}\sim1.1 \times \sqrt{2}$ as shown in Figure~\ref{fig2} (a). It is different from the case when $v_{0}\ll\sigma$ because of the contribution of the higher order terms in $v_{0}$. Therefore, we get $\sigma_{nt}/v_{rms}\sim1.1$ instead of 1 (when $v_{0}\ll\sigma$). Intuitively, one can say that the period average spectrum is computed by averaging the spectra that are equally shifted in blue-ward and red-ward directions by $v_{0}$. Thus, the nonthermal width of the period average spectrum is equal to $v_{0}$. 
These relations differ from those used in the earlier studies where either $\sigma_{nt}/v_{rms}=1$ or $\sigma_{nt}/v_{rms}=1/\sqrt{2}$.\\
The scenario we discuss in this section is too ideal to be realistic. The exposure time of the spectrographs such as Extreme Ultraviolet Imaging Spectrometer (EIS) onboard Hinode and Coronal Multichannel Polaimeter (CoMP) have cadence less than 60 s which is less than the characteristic transverse wave period ($\sim$5 min). In later sections we discuss more general scenarios, where we include the inclination of LOS with oscillations, superposition of different of polarisation of oscillations and superposition of many wave modes and relax the requirement of a large exposure time.


\subsection{LOS inclined to the oscillations' direction}
If the direction of oscillations are inclined to the LOS at an angle $\theta$ (Figure~\ref{fig1} (a)), the time-integrated spectrum is given by the following equation,
\label{sec2.2}
\begin{equation}
\begin{aligned}
    \big \langle G(\mathfrak{v},t') \big \rangle _{t}=\int_{0}^{P} \frac{1}{P \sigma \sqrt{2 \pi}} \exp(-\frac{(\mathfrak{v}-v_{0}~\cos(\theta)~\cos(\omega t'))^{2}}{2\sigma^{2}}) dt'.\\
\end{aligned}
\label{eq4}
\end{equation}
Note that equation~\ref{eq4} is similar to Equation~\ref{eq2}, except that the wave amplitude is modulated along the LOS. Borrowing the relation from the previous scenario, we get $\sigma_{nt} \sim\sqrt{2}~v_{rms}~\cos\theta$ $\sim$ $v_{0}~\cos\theta$ when $v_{0}\ll\sigma$. Taking $v_{0}$~=11$\sqrt{2}$~km~$s^{-1}$ and $\sigma$=19/$\sqrt{2}$~km~$s^{-1}$, we get $\sigma_{nt} \sim 1.1~\sqrt{2}~v_{rms}~\cos\theta$ $\sim$ 1.1 $v_{0}~\cos\theta$.
\subsection{Effect of LOS superposition of oscillating structures on the nonthermal line widths}
\label{sec2.3}
Now, we assume that different structures are positioned along the same LOS but oscillating in different directions. At a given time $t_{0}$, the emitted spectra from the oscillating structures along the LOS will superimpose to generate a broader spectrum. Thus the nonthermal line widths will increase due to unresolved wave amplitudes along the LOS. The shape of the normalised spectrum at an instant ($t_{0}$) is given by the following relation,
\begin{equation}
\begin{aligned}
    \big \langle G(\mathfrak{v},\theta ') \big \rangle _{\theta}=\int_{0}^{2\pi} \frac{1}{2\pi \sigma \sqrt{2 \pi}} \exp(-\frac{(\mathfrak{v}-v_{0}\cos(\theta ')~\cos(\omega t_{0}))^{2}}{2\sigma^{2}}) d\theta '.
\end{aligned}
\label{eq5}
\end{equation}
Here wave amplitude ($v_{0}$) is modulated depending on the angle between the direction of oscillations and LOS. Note that Equation~\ref{eq5} is similar to Equation~\ref{eq2} because $t_{0}$ is a fixed constant. Therefore in this case too, $\sigma_{nt}\sim\sqrt{2}v_{rms}$=$v_{0} \cos(\omega t_{0})$ for the limiting case ($v_{0}\ll\sigma$). Otherwise, taking $v_{0}$~=11$\sqrt{2}$~km~$s^{-1}$ and $\sigma$=19/$\sqrt{2}$~km~$s^{-1}$, we find $\sigma_{nt}\sim1.1\sqrt{2}  v_{rms}\sim 1.1~v_{0} \cos(\omega t_{0})$. Depending on the phase of oscillation ($t_{0}$), the nonthermal line width will change. 
Here, we make a rigid assumption that all structures along the LOS are oscillating in (or out of)-phase because $t_{0}$ is assumed constant. No other value of phase difference is possible. This may not be a realistic scenario but for the sake of completeness, we discuss it here.

\subsection{Effect of different polarizations and phase of oscillations on nonthermal line widths}
\label{sec2.4}
Perhaps the most probable scenario in the optically thin corona is the occurrence of structures oscillating in different directions and different phases superposed along the LOS. Assuming a uniform probability of choosing oscillating structures in a given state of polarisation and phase of oscillation, the time-integrated spectrum over one period can be computed by the following relation,
\begin{equation}
\begin{aligned}
    \big \langle G(\mathfrak{v}) \big \rangle _{t,\theta}=\frac{1}{P \sigma (2 \pi)^{3/2}}\int_{0}^{2\pi} \int_{0}^{P} \exp(-\frac{(\mathfrak{v}-v_{0}\cos\theta' \cos(\omega t'))^{2}}{2\sigma^{2}}) dt' d\theta'.
\end{aligned}
\label{eq6}
\end{equation}
For simplicity we are integrating up to one period of oscillation only. We also do not assume any constraint on the exposure time of the measuring instrument. It should be noted that as long as different phases and polarizations are equally likely and total number of structures oscillating along the LOS is large, Equation~\ref{eq6} will always be applicable within reasonable accuracy despite the exposure time of the measuring instrument is less than the wave period.\\
We solve Equation~\ref{eq6} analytically assuming $\theta'$ and $t'$ are independent of each other and $v_{0}\ll\sigma$ (see appendix~\ref{appb}). We first integrate equation~\ref{eq6} in time to obtain an expression described by the equation~\ref{eqq11} for every $\theta'$. Thus for a given $\theta'$, the nonthermal width of a period averaged spectrum is equal to $v_{0}\cos\theta'$. Next, we average over all directions with LOS ($\theta'$) to obtain $\big \langle G(\mathfrak{v}) \big \rangle _{t,\theta}$. Intuitively, this is somewhat similar to the rms averaging of spectra of different nonthermal widths. Thus, the nonthermal line widths in this scenario is further reduced by $\sqrt{2}$ compared to the scenario described in section~\ref{sec2.1}. Therefore, we find that $\sigma_{nt}/v_{rms}\sim1$.
Next, we relax the assumption $v_{0}\ll\sigma$ and choose $v_{0}$~=11$\sqrt{2}$~km~$s^{-1}$ and $\sigma$=19/$\sqrt{2}$~km~$s^{-1}$.
We numerically integrate Equation~\ref{eq6} using the quadrature method\footnote{\url{https://docs.scipy.org/doc/scipy/reference/generated/scipy.integrate.dblquad.html}} and compute the normalized spectrum, which is shown in Figure~\ref{fig2} (b).  We compute the nonthermal line widths and estimate that $\sigma_{nt}/v_{rms}\sim$1.1. Note that it is less than those described in above sections where $\sigma_{nt}/v_{rms}\sim$1.1~$\sqrt{2}$.

Until now we have assumed only linearly polarised transverse oscillations. Now, we assume circularly polarised oscillations \citep{1955RSPSA.233..310F}. A circularly polarised oscillations can be expressed mathematically as ${\bf v}=\frac{1}{2}[v_{0}e^{-i\omega t} {\bf e_{+}} + v^{*}_{0}e^{i\omega t} {\bf e_{-}}]$, where ${\bf e_{\pm}}=\frac{1}{\sqrt{2}}({\bf e_{x} \pm i e_{y}})$ \citep{1978ApJ...219..700G}. Taking only real velocity amplitudes, we get ${\bf v}=\frac{v_{0}}{\sqrt{2}}[\cos(\omega t) {\bf e_{x}} + \sin(\omega t) {\bf e_{y}}]$. Thus, for a circularly polarised oscillation, polarisation direction along two perpendicular axes has a phase difference of $\pi/2$ shown schematically in Figure~\ref{fig1}. 
The time-integrated spectrum due to such oscillations can be computed by the following expression,
\begin{equation}
\big \langle G(\mathfrak{v}) \big \rangle _{t,\theta}=\frac{1}{P \sigma (2 \pi)^{3/2}}\int_{0}^{2\pi} \int_{0}^{P} \exp(-\frac{(\mathfrak{v}-v_{01}\cos(\theta'-\omega t'))^{2}}{2\sigma^{2}}) dt' d\theta'.
\label{eq7}
\end{equation}
Here $v_{01}=\frac{v_{0}}{\sqrt{2}}$. We use $v_{01}$ instead of $v_{0}$ for consistency such that the energy in the linear and circularly polarised driver remains equal.
First, note that the integration along $\theta$ is redundant because a circularly polarised oscillation will cycle through all angles relative to the LOS. Therefore equation~\ref{eq7} is equivalent to equation~\ref{eq2}. Thus for a circularly polarised velocity driver as described above, $\sigma_{nt}\sim v_{01}$ for the limiting case and $\sigma_{nt}\sim~1.1v_{01}$, when $v_{01}$~=11~km~$s^{-1}$ and $\sigma$=19/$\sqrt{2}$~km~$s^{-1}$ as shown in Figure~\ref{fig2} (b).\\
It is worth combining equations for different velocity drivers described above into a single equation.  Equations~\ref{eq6} and~\ref{eq7} describe the shape of the period averaged spectrum for a linearly and circularly polarised driver. To combine these two equations, we define a new variable $\phi$ which may or may not be a function of time depending on the nature of the velocity driver. Mathematically, $\phi(t)$ is defined as follows. 
\begin{equation}
\phi(t)=\tan^{-1} \big(\frac{v_{y} \sin (\omega_{y}t + \delta)}{v_{x} \sin (\omega_{x}t)}\big),
\label{phi}
\end{equation}
where $\delta$ is the phase difference between $x$ and $y$ components of the velocity driver. $\omega_{x}$ and $\omega_{y}$ are the frequencies of oscillations of the velocity driver in $x$ and $y$ directions. From equation~\ref{phi}, we note that for a linearly polarised driver with equal amplitudes and frequency in $x$ and $y$ axes, $\delta$ vanishes. Thus $\phi=\pi/4$ and constant in time. However, for a circularly ($v_{x}=v_{y}$ and $\omega_{y}=\omega_{x}$) polarised driver, $\delta=\pi/2$. Thus, $\phi=\omega t$. $\phi$ can be considered as the angle that velocity driver make with $\bf{e_{x}}$. $\theta$ used in the equations above is the angle that LOS make with the $\bf{e_{x}}$. Thus the effective angle between LOS and direction of oscillation at an instant can be taken as $\theta - \phi(t)$. \\
Also, we define total velocity of the driver as follows.
\begin{equation}
    v_{T}(t)=\sqrt{(v_{x} \sin (\omega_{x}t))^{2} + (v_{y} \sin (\omega_{y}t + \delta))^{2}}
\label{tv}
\end{equation}
Expressions of total velocity $v_{T}(t)$ and $\phi(t)$ can be used to obtain period averaged spectrum can be computed by the following expression. 
\begin{equation}
\big \langle G(\mathfrak{v}) \big \rangle _{t,\theta}=\frac{1}{P \sigma (2 \pi)^{3/2}}\int_{0}^{2\pi} \int_{0}^{P} \exp(-\frac{(\mathfrak{v}-v_{r}(t')\cos(\theta'-\phi(t')))^{2}}{2\sigma^{2}}) dt' d\theta'.
\label{eq8}
\end{equation}

Note that under the assumption {\bf $v_{x}=v_{y}=\frac{v_{0}}{\sqrt{2}}$}, $\omega_{x}=\omega_{y}=\omega$; for $\delta=0$ and $\pi/2$, equation~\ref{eq8} reduces to equations~\ref{eq6} and~\ref{eq7}, respectively.\\ 
Thus we conclude that for the limiting case when $v_{0}\ll\sigma$, $\sigma_{nt}/v_{rms}\sim1$. Taking wave amplitudes ($v_{0}$) of 11$\sqrt{2}$~km~$s^{-1}$ and $\sigma$=19/$\sqrt{2}$~km~$s^{-1}$ in the solar corona, $\sigma_{nt}/v_{rms}\sim1.1$ for a linearly and circularly polarised oscillation when averaged over period and direction of oscillations.
Equations~\ref{eq8} is quite useful for deriving the relation between the nonthermal line widths and rms wave velocities if multiple wave drivers are assumed or if a random distribution is assumed instead of a uniform distribution. We again reiterate that the wave amplitude is assumed to be smaller than the thermal line width of the optically thin emission line. The derived relations may not be valid if wave amplitudes are larger than thermal line widths. Figure~\ref{fig2} (d) shows the period averaged spectrum (blue curve) for $v_{rms}=22~km~s^{-1}$ when line width, $\sigma$ is assumed to be 19$/\sqrt{2}~km~s^{-1}$. It can be seen for a large amplitude, the shape of the averaged spectrum is no longer a Gaussian. 


\subsection{Multiple wave drivers}
\label{sec2.5}
\begin{figure}[!ht]
\centering
\includegraphics[scale=0.5]{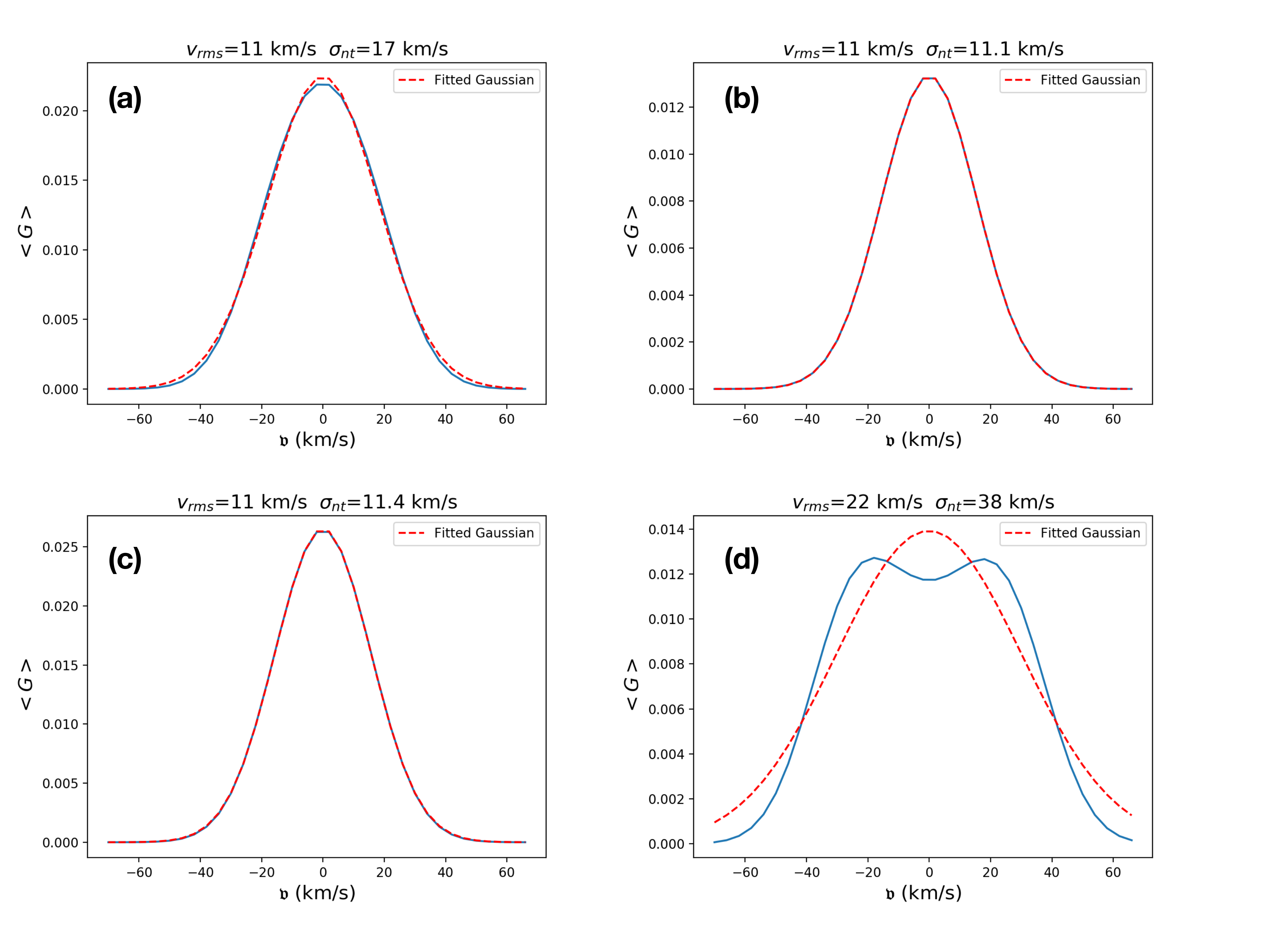}
\caption{Best-fit Gaussian curves in red over the integrated spectra shown in blue for the (a) linear, (b) circular, and (c) multiple velocity drivers. Note that $\frac{\sigma_{nt}}{v_{rms}}$ $\sim$ 1.5, 1, 1 respectively for these three scenarios. (d) Averaged spectrum when the strength of the velocity driver is greater than the thermal line width.}
\label{fig2}
\end{figure}

Here, we investigate the effect of the multiple wave drivers on the rms wave velocities and nonthermal line widths. Such drivers are used in the earlier studies \citep{2017NatSR...714820M,2019ApJ...881...95P} to excite waves in 3D MHD simulations. We assume a superposition of ten different velocity drivers with different velocity amplitudes and periodicity given by the following relations.
\begin{equation}
\begin{aligned}
v_{x}(t)=\sum_{i=1}^{10} v_{xi} \sin(\omega_{xi} t),\\
v_{y}(t)=\sum_{i=1}^{10} v_{yi} \sin(\omega_{yi} t).\\
\label{eq10}
\end{aligned}
\end{equation}
Using equations~\ref{phi} and \ref{tv}, We compute $v_{T}=\sqrt{v_{x}^{2} + v_{y}^{2}}$ and $\phi (t)=\tan^{-1}(v_{y}/v_{x})$. Note that the sum over different velocity drivers is computed at every instant of time. Also the amplitude and frequency of ten drivers do not change with time. We insert $v_{T}$ and $\phi$ in Equation~\ref{eq8} to compute the period averaged spectrum shown in Figure~\ref{fig2} (c) in blue. Fitting a Gaussian function over the numerically integrated period averaged spectrum, we find that $\sigma_{nt}/v_{rms}\sim$1.1.


We learn from these ideal cases that when different structures with different polarisations are oscillating in different phases, the nonthermal line widths are at least equal to the rms wave velocities. Further we note that when the wave amplitude is of the order of the thermal line widths, $\sigma_{nt}/v_{rms}\sim1.1$. These results are in contrast with those used in the earlier studies by \citet{1990ApJ...348L..77H,b98,2005A&A...436L..35O,b2009,2012ApJ...753...36H} where $\frac{\sigma_{nt}}{v_{rms}}$ $\sim$ 1/$\sqrt{2}$=0.71 is used. Thus we find that the Alfv\'en(ic) wave energy flux was overestimated by at least a factor of two in these studies.


\section {Numerical simulations and forward modeling}
\label{sec3}
\begin{figure}
\centering
\includegraphics[scale=0.5]{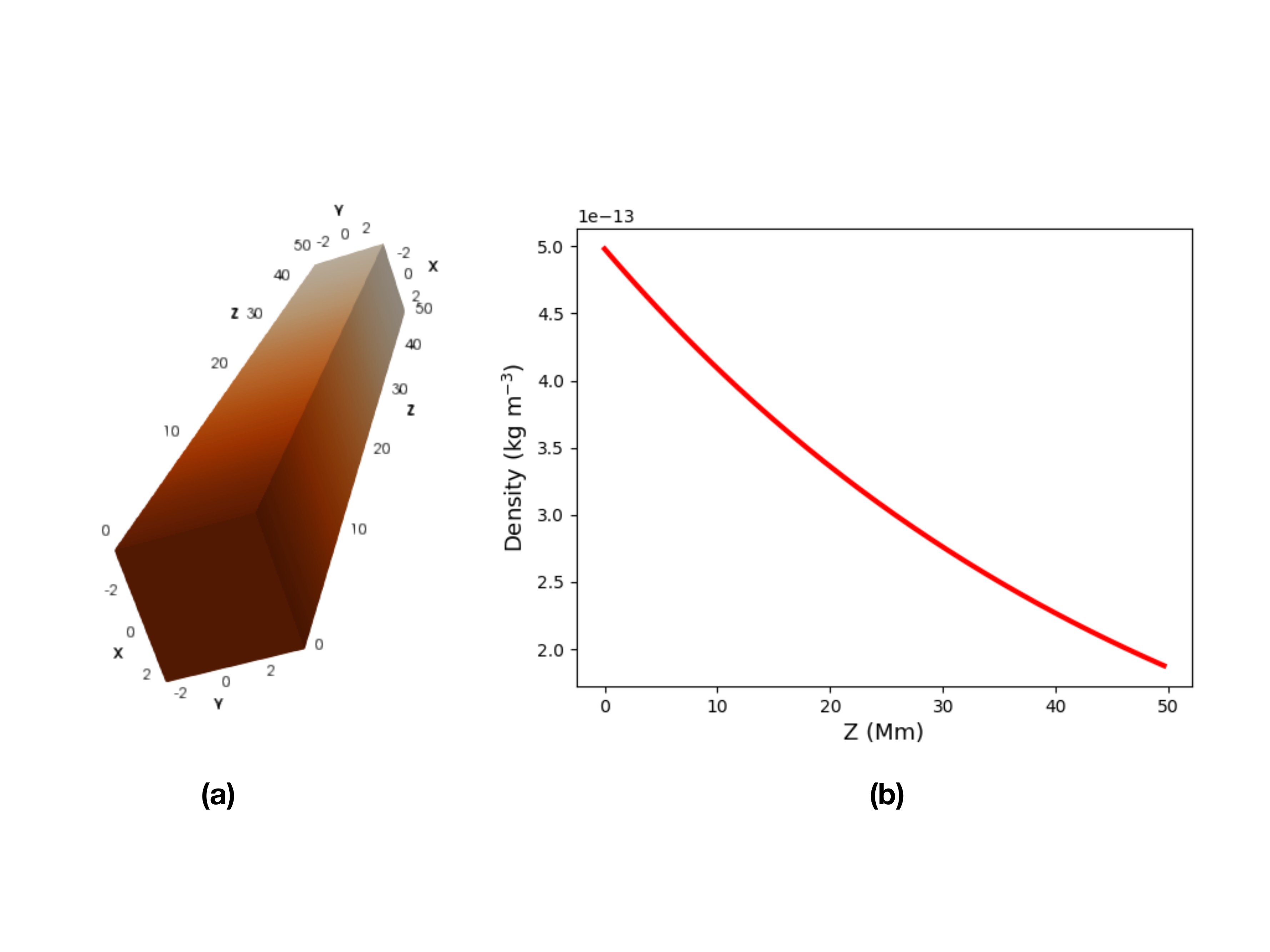}
\caption{(a) Initial simulation set-up. Density (in red) is stratified along the $z$-direction. (b) Variation of density averaged over $x$ and $y$ directions along height. The density is uniform across the background magnetic field. }
\label{fig3}
\end{figure}

We test the validity of the mathematical models described in Section~\ref{sec2} using physical models employing ideal 3D MHD simulations using MPI-AMRVAC that solves the following equations in the near conservative form \citep{2014ApJS..214....4P}. 

\begin{equation}
\begin{aligned}
\frac{\partial \rho}{\partial t} + {\bf \nabla} . (\rho {\bf v})=0, \\
\frac{\partial ({\rho {\bf v}})}{\partial t} + \nabla.(\rho {\bf v v} - {\bf BB}) + \nabla (p + {\bf B}^{2}/2) - \rho {\bf g} = 0,\\
\frac{\partial E}{\partial t} + \nabla.({\bf v}E - {\bf BB.v} + {\bf v}.(p + {\bf B}^{2}/2)) = 0,\\
\frac{\partial B}{\partial t} - \nabla \times ({\bf v} \times {\bf B}) = 0,\\
\nabla.{\bf B} = 0.
\end{aligned}
\label{eq11}
\end{equation}
Here {\bf g} is the acceleration due to gravity of the sun pointing along negative $z$ axis, $E$ is total energy density defined as, $E=\frac{p}{\gamma -1} + \frac{\rho {\bf v}^{2}}{2} + \frac{{\bf B}^{2}}{2}$. $\rho$ is the density which is an exponentially decaying function of the height ($x$ axis) due to the gravitational stratification. We choose a background magnetic field strength ($B$) of 5 G along $z$ axis in all simulation runs.
We perform numerical simulations for transversely homogeneous plasma for mono-periodic and multi-periodic velocity drivers. We choose a transversely homogeneous plasma because we want to study the effects of LOS superposition without generating the uniturbulence due to the transverse inhomogeneity in the density.\\
The set of equations described in Equation~\ref{eq11} are solved in the Cartesian geometry for a grid size of 64$\times$64$\times$128 that span a physical dimension of 5 Mm$\times$5 Mm $\times$ 50 Mm as shown in Figure~\ref{fig3}. The geometry and size of the simulation set-up is similar to the one described in \citet{2019ApJ...881...95P}. Since we assume transversely homogeneous plasma, we used a coarse grid resolution along the $x$ and $y$ directions. The side boundaries of the simulation domain are periodic while the top boundary is kept open so that waves can leave the simulation domain. Plasma beta ($\beta$) and temperature of the plasma are kept at 0.07 and 1.2 MK, respectively, at the start of the simulation. We let simulations reach a quasi-steady state before implementing the velocity drivers at the bottom boundary. In the quasi-steady state, the density varies exponentially (Figure~\ref{fig3} (b)) and the scale height was estimated to be $\sim$ 51 Mm. Figure~\ref{fig3} (a) shows the density distribution in the initial configuration of the simulation cube. Figure~\ref{fig3} (b) shows the variation of density with height ($z$ axis).\\
First, we excite the bottom boundary of the simulation with a linearly polarised transverse velocity driver given by the following relations.
\begin{equation}
\begin{aligned}
v_{x}(z=0,t)=U_{0} \sin(\omega_{0} t),\\
v_{y}(z=0,t)=V_{0} \sin(\omega_{0} t),\\
\end{aligned}
\label{eq12}
\end{equation}
We choose $U_{0}$=$V_{0}$=11 $km~s^{-1}$ and the period of oscillations, $P=2\pi/\omega_{0}$=400 s.\\

\begin{figure}
\centering
\includegraphics[scale=0.5]{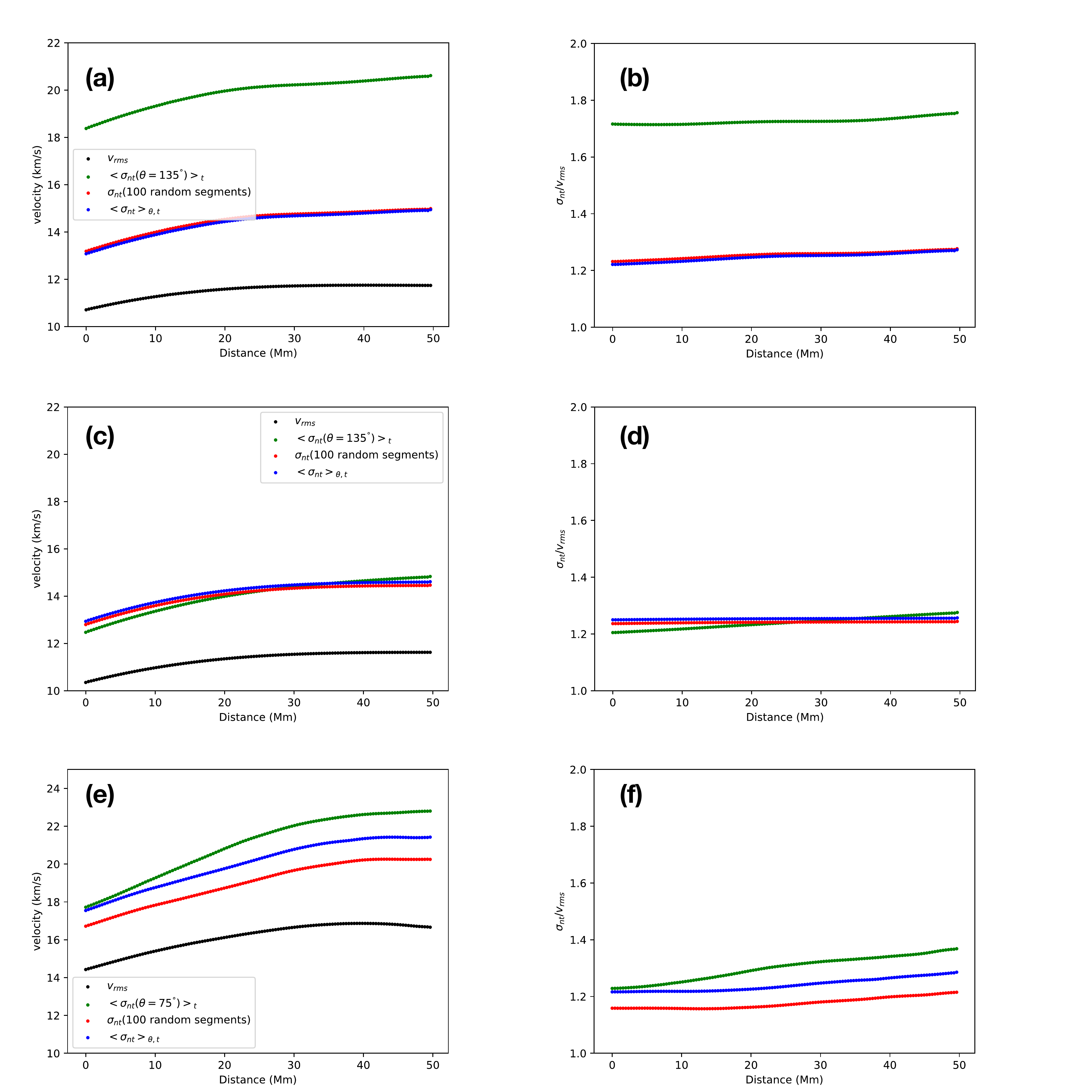}
\caption{Variation of $v_{rms}$ and $\sigma_{nt}$ with height for different scenarios. $v_{rms}$ computed from simulation is shown in black. $\sigma_{nt}$ computed from a period averaged spectrum for a given LOS is shown in green. $\sigma_{nt}$ for a period (and LOS) averaged spectrum and spectrum averaged over 100 random segments are overplotted in blue and red. (a), (c), and (e) panels represent the results for linear, circular and multi-frequency velocity driver.
(b), (d), and (f) present the variation of the ratio of $\sigma_{nt}$ and $v_{rms}$ for linear, circular and multi-frequency velocity driver.}
\label{fig4}
\end{figure}

To study the relationship between the nonthermal line widths and rms wave velocities, we perform forward modeling of the simulations for 12 different LOS using the Fe XIII emission line (10749 \AA\ ) in FoMo-c\footnote{\url{https://wiki.esat.kuleuven.be/FoMo/FoMo-C}} \citep{10.3389/fspas.2016.00004}. We adopted the same methodology as described in \citet{2019ApJ...881...95P} for estimating the nonthermal line widths at all LOSs. Using simulated data, we compute the variation of the rms wave velocity with height (shown in black in Figure~\ref{fig4}). Further, we compute the nonthermal line widths for a LOS inclined at an angle $\theta$ to the direction of oscillations (shown in green in Figure~\ref{fig4}). We also estimate the nonthermal line widths after averaging over period and direction of oscillations (curves in blue in Figure~\ref{fig4}). Finally, we choose 100 random segments with random direction of oscillations and phases and compute the nonthermal line widths using the spectrum averaged over these segments (overplotted in red in Figure~\ref{fig4}). The right hand panels of Figure~\ref{fig4} show the ratio of the nonthermal line widths and rms wave amplitude with height. In Figure~\ref{fig4} (a) and (b), we note that $\sigma_{nt}/v_{rms} \sim$ 1.2$\sqrt{2}$ when the LOS is aligned in the direction of the oscillations which is 135$^{\circ}$ and the spectra are integrated over time. This relation matches fairly well with those predicted using the numerical integration in section ~\ref{sec2.1}. Similarly,  $\sigma_{nt}/v_{rms} \sim$ 1.2 when spectra are integrated spatially and temporally as shown in blue. It should be noted that the results for the random segments (overplotted in red) are similar to those obtained assuming a uniform probability of occurrence of different polarisation and phases of the oscillation along LOS. The small deviation from the blue could be due to the randomness in choosing segments with different phase and polarisation of oscillations. Again, these results match with those discussed in section~\ref{sec2}.\\
Next, we implement the velocity drivers with a phase difference of $\pi$/2 in $y$ and $z$ directions leading to circularly polarised oscillation.

\begin{equation}
\begin{aligned}
v_{x}(z=0,t)=U_{0} \sin(\omega_{0} t),\\
v_{y}(z=0,t)=V_{0} \cos(\omega_{0} t).\\
\end{aligned}
\end{equation}
The period and velocity amplitude of oscillations are similar to those described above. Figure~\ref{fig4} (c) and (d) show the results of this simulation run. We find that irrespective of the LOS chosen, $\sigma_{nt}/v_{rms} \sim$ 1.2, which is in good agreement with those derived in section 2 for the circularly polarised oscillations.

Finally, we implement the multiple (ten) velocity drivers described by the Equation~\ref{eq13} at the bottom boundary \citep[see][]{2017NatSR...714820M,2019ApJ...881...95P} and repeat the analysis described above:
\begin{equation}
\begin{aligned}
v_{x}(z=0,t)=\sum_{i=1}^{10} U_{i} \sin(\omega_{i} t),\\
v_{y}(z=0,t)=\sum_{i=1}^{10} V_{i} \sin(\omega_{i} t).\\
\end{aligned}
\label{eq13}
\end{equation}
We choose $\omega_{i}$ from the observed distribution of transverse oscillation period in the coronal holes \citep{2015NatCo...6E7813M}. $U_{i}$ and $V_{i}$ are randomly chosen such that $v_{rms}$ at the bottom boundary $\sim$ 16 km s$^{-1}$ \citep[see][for details]{2019ApJ...881...95P}. Figures~\ref{fig4} (e) and (f) present the results for this velocity driver. We notice that $\sigma_{nt}/v_{rms}$ $\sim$ 1.2 for the scenarios when averaging over 100 random segements and over period and LOSs are performed. These results match with those presented in the section ~\ref{sec2.5}. It is worth noting that for $\theta=75^{\circ}$, $\sigma_{nt}/v_{rms}$ is between 1.2 and 1.2$\sqrt{2}$. In fact, we note that period averaged spectrum for any LOS is $1.2 \le \sigma_{nt}/v_{rms} \le 1.2\sqrt{2}$. This happens because such multiple velocity drivers result in the velocity field which is neither circular nor linear but forms a Lissajous patterns \citep[see online animation in][]{2019ApJ...881...95P}. It is worth noting that in all the above described scenarios $v_{rms}$ is less than the thermal width of the emission line, which is 19 km s$^{-1}$.\\
Key results obtained from this study are that for a transversely homogeneous and gravitationally stratified plasma $\sigma_{nt}/v_{rms}$ $\sim$ 1.2 for scenario when different polarisation and phase of oscillations occur along the LOS of an observer. We use both mathematical model and 3D MHD simulations together with forward modeling to verify these relations. These relations are valid for both single- and multi-frequency velocity drivers.

\section{Wave amplitudes {\it vs} height}
\begin{figure}
\centering
\includegraphics[scale=0.8]{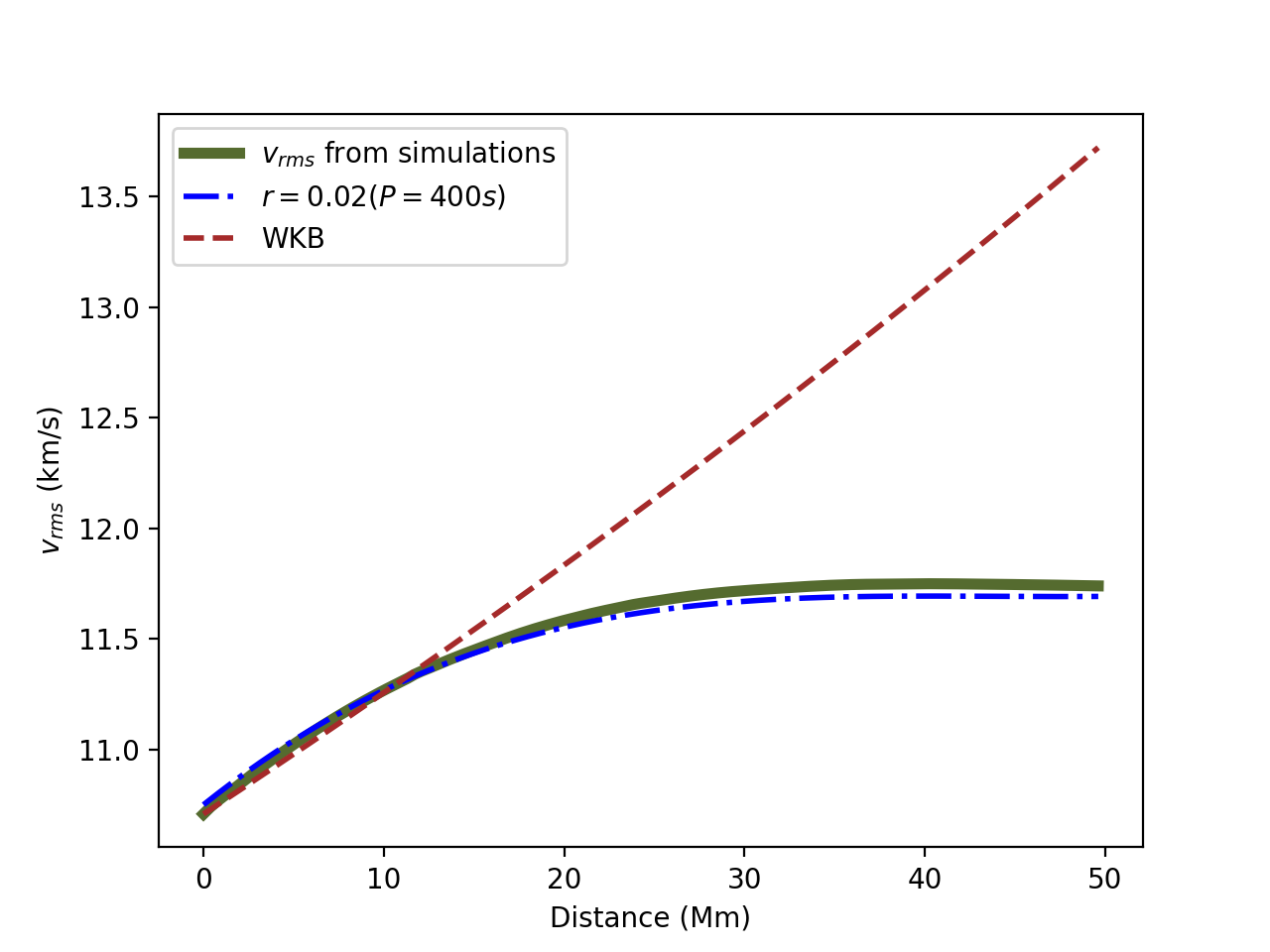}
\caption{Variation of the rms wave velocity with height computed from the simulations. Overplotted in blue is the best-fit curve obtained by using equation~\ref{eq16} for $r=0.02$. $v_{rms}$ estimated using a WKB approximation is overplotted in dark red.}
\label{fig5}
\end{figure}

We note that regardless of the nature (linear or circular) of the polarisation of the transverse MHD waves, the wave amplitudes and hence line widths increase and level-off with height in our simulations. This behavior is noted for both transversely homogeneous (this study) and inhomogeneous simulations \citep[see][]{2019ApJ...881...95P}. In this section we try to understand this nature of variation.\\
We eliminate the possibility of damping due to resonant absorption or numerical dissipation by performing the 3D MHD simulations of a homogeneous plasma without gravity. We did not find any significant damping of the wave amplitudes. Wave amplitudes are reduced by $\sim$ 2-3\%. This means wave energy is reduced by 4\%.
In observations, it is often assumed that the rms wave amplitude of Alfv\'en(ic) waves increases with height assuming a Wentzel-Kramers-Brillouin (WKB) approximation. Under this approximation $v_{rms}$ varies as $\rho^{-1/4}$ \citep{1972ApJ...177..255H}. Using this expression, $v_{rms}$ at the base of the corona is estimated by comparing it with the observed values of the nonthermal line widths \citep{1990ApJ...348L..77H,b98}. However, the line widths appear to increase more slowly than expected from WKB propagation (level-off) at higher heights \citep{1990ApJ...348L..77H,b98,2012ApJ...753...36H}. The deviation from WKB propagation is considered to be the signature of the damping of waves \citep{2012ApJ...753...36H}. In our simulations, the damping of wave energy is insignificant, this means the wave amplitude should increase according to the WKB approximation and should be around $\sim$ 13.7 km $s^{-1}$ at 50 Mm (dashed curve in red in Figure~\ref{fig5}). However, we note from Figure~\ref{fig4} (a), the wave amplitudes are around 11.9 km s$^{-1}$. This corresponds to a 15\% difference in the wave amplitudes between simulations and those expected from the WKB theory.\\
Also, note that the wavelength of the transverse oscillation is $\sim$ 300 Mm in our simulations. Since the wavelength is much larger than the scale height of the simulations (50 Mm), the WKB approximation may not be valid in this scenario \citep{1972ApJ...177..255H,1978SoPh...56..305H,1980JGR....85.1311H}.\\
To understand the propagation of transverse MHD waves in gravitationally stratified medium, we use the expressions derived by \citet{1978SoPh...56..305H} for propagating transverse waves without assuming WKB (or eikonal) approximation. We borrow the following relation from  \citet{1978SoPh...56..305H}.
\begin{equation}
    v=a H_{0}^{(1)} (\alpha) + b H_{0}^{(2)} (\alpha).
    \label{eq14}
\end{equation}
Here, $v$ is the velocity amplitude, $H_{0}^{(1)}$ and $H_{0}^{(2)}$ are the Hankel functions of the first and second kinds respectively, and $\alpha=\frac{2H\omega}{v_{A}(z)}$. $H$ is the scale height, $\omega$ is the frequency and $v_{A}(z)$ is the Alfv\'en velocity which is a function of height ($x$-direction). $a$ and $b$ are unknowns to be derived from the boundary conditions.\\
 The reflection coefficient, $r$, is defined as, $r=\frac{|b|^{2}}{|a|^{2}}$. $r$=0 means no reflection (only outward propagating waves).
Using  equation~\ref{eq14}, one can show that \citep[see][]{1978SoPh...56..305H} 
\begin{equation}
    Re(v)=R_{2} cos(\omega t) - I_{2} sin(\omega t).
    \label{eq15}
\end{equation}
This equation leads to the estimation of the rms velocity which is given by
\begin{equation}
    v_{rms}=\sqrt{(R_{2}^{2} + I_{2}^{2})/2}.
    \label{eq16}
\end{equation}
Here,
\begin{equation}
\begin{aligned}
    R_{2}=Re(a)J_{0} - Im(a) Y_{0} + Re(b) J_{0} + Im(b)Y_{0},\\
    I_{2}=Re(a)Y_{0} + Im(a) J_{0} - Re(b) Y_{0} + Im(b)J_{0},
    \end{aligned}
    \label{eq17}
\end{equation}
where, J and Y are Bessel functions.\\
Using equation~\ref{eq16}, we computed the variation of $v_{rms}$ with height by iterating over several values of complex numbers $a$ and $b$ and performing chi-square minimisation over rms wave amplitudes computed from simulations. We choose those values for which the analytical expression given by Equation~\ref{eq16} matches fairly well with those obtained from simulations. The best-fit curve in blue obtained using equation~\ref{eq16} is overplotted on the rms wave amplitude for the linearly polarised velocity driver overplotted with green in Figure~\ref{fig5}. We find a fairly good match between the two curves. Finally, we estimate the reflection coefficient, $r$, which was found to be 2.5\% in the case of linearly polarised oscillations. For the circularly polarised and multi-frequency drivers, the reflection co-efficient was found to be 1.5-3\%. It is worth noting here that a reflection coefficient of 2.5\% leads to a difference of 15\% in the wave amplitudes. This matches with the difference in the wave amplitudes between simulations and those expected from WKB.\\
We also note that when wavelength of the wave is much less than the scale height, the rms wave velocity will match fairly with those computed assuming WKB approximation. 
We notice that the deviation from the expected WKB propagation in our simulations could be due to the presence of incoming wave ($b \ne 0$) due to gravitational stratification leading to a non-WKB effects. 
Also, we note that the wave amplitude are smaller than those expected from WKB approximation. This is consistent with the study of \citet{1972ApJ...177..255H}, where authors reported a similar nature of variation \citep[see figure~1 and figure~7 of][respectively]{1972ApJ...177..255H,1981SoPh...70...25H}. Thus both large wavelength and reflection affects the observed wave amplitudes and hence line widths in the solar atmosphere.\\
This leads us to believe that the reflection coefficient as low as 2\% can significantly change the nature of the variation of wave amplitudes with height. Thus a slower increase of the wave amplitudes and hence nonthermal line widths than expected from the WKB approximation is not only the signatures of the damping but could also be due to the non-WKB nature of transverse wave propagating in a gravitational stratified plasma. In the polar coronal holes, the departure from the WKB theory appears to happen at heights of 0.2 R$_{\sun}$ (from the photosphere) or ~140 Mm \citep{b98,2012ApJ...753...36H}. In this study, the effect is seen after 10 Mm. There might be several reasons for it. First, our simulations assume an isothermal atmosphere with a scale height of $\sim$50 Mm. The scale heights in the solar corona might be very large. For example, \citet{b98}, reported a scale height of 100 Mm \citep[see Figure 4a in][]{b98}. Similarly, \citet{2008A&A...483..271D} reported a scale height of $\sim$70 Mm. Recently, \citet{2019ApJ...884...43P} investigated the density and temperature variation along height in the coronal holes. They found that the density scale height changes with height. We have not considered these effects in our simulations. These might affect the rms velocities and nonthermal line widths. Second, the nature of the variation of the nonthermal line widths is different in different emission lines \citep{2008A&A...483..271D}. Third, the observed nonthermal line widths, rather than wave amplitudes, are compared with the WKB calculations. Though the nonthermal line widths depend on the rms wave amplitudes, in a realistic atmosphere, there will be other effects such as turbulence that can change the nature of the variation of the nonthermal line widths with height. For example, \citet{2019ApJ...881...95P} reported that the nonthermal line widths did not level-off significantly with height when $v_{rms}=26~ km~s^{-1}$ as compared to the scenarios when $v_{rms}=11$ and $7~km~s^{-1}$ \citep[see Figure~10 in][]{2019ApJ...881...95P}. Fourth, the mismatch between $v_{rms}$ and those obtained using WKB approximation might also depend on the nature of the velocity driver. For example, the nature of the variation of the nonthermal line widths of random segments for a multifrequency driver is slightly different from the nature of the variation of rms wave amplitude (see Figure~\ref{fig4}(e)). This difference might be due to the choice of the random segments. Finally, it should be noted that the difference in the wave amplitudes computed using WKB theory and the real wave amplitudes in Figure~\ref{fig5} is of the order of 1-2 km~s$^{-1}$. In real observations, this difference might appear within the error bars of the measurements. Recently, \citet{2020ApJ...894...79W} compute the transverse wave amplitudes using Atmospheric Imaging Assembly (AIA)/ Solar Dynamics Observatory (SDO) and note a similar variation of the wave amplitudes with height. In their study, the flattening of wave amplitudes happens at 15 Mm. Interestingly, these authors also suggest the possibility of the reflection in the low corona. However, they do not attribute the flattening of wave amplitudes to the reflection. Thus, we believe that wave amplitudes, rather than the nonthermal line widths, are more meaningful to compare.

\begin{table}[htb]
\centering
\caption{Ratio of nonthermal line width and rms wave amplitude for different scenarios assuming $v_{0}=$11$\sqrt{2}$~km~$s^{-1}$ and $\sigma$=19/$\sqrt{2}$~km~$s^{-1}$.}  
\label{table1}
\scalebox{0.9}{
\begin{tabular}{|c|c||c|c|}
 \hline
\multicolumn{2}{|c||}{Nature of velocity driver} & Theoretical $\sigma_{nt}/v_{rms}$ & Numerical $\sigma_{nt}/v_{rms}$\\
\hline
\multirow{2}{*}{Mono-periodic linearly polarised} & oscillations along LOS & $1.1\sqrt{2}$ & $1.2\sqrt{2}$\\
& oscillations superimposed along LOS & $1.1$ & $1.2$\\
\hline
\multicolumn{2}{|c||}{Mono-periodic circularly polarised} & $1.1$ & $1.2$\\
\multicolumn{2}{|c||}{Multi-frequency driver} & $1.1$ & $1.2$\\
\hline
\end{tabular}
}
\label{table1}
\end{table}

\section{Summary and Conclusions}
\citet{2012ApJ...761..138M,2012ApJ...746...31D,2019ApJ...881...95P} have shown that LOS superposition of oscillating structures greatly reduce the rms Doppler velocities and increase nonthermal line widths. Thus real wave amplitudes are hidden in the nonthermal line widths. This motivated us to estimate a relation between nonthermal line widths and rms wave velocities which can be used to estimate the true energy carried by the Alfv\'en(ic) wave. Furthermore, estimating wave amplitudes using the nonthermal line widths is useful because if high-resolution spectral and imaging data is available, the resolved wave amplitudes either in the plane-of-sky; POS or along LOS (exhibiting the periodic variation in the Doppler velocities) can be combined to compute true wave amplitudes and hence energies.\\
We build a mathematical model and found the relation between $\sigma_{nt}$ and $v_{rms}$ for different scenarios. The results are summarised in Table~\ref{table1}.  For the limiting case, when $v_{0}\ll\sigma$ and if a single structure is assumed to oscillate along the LOS, $\sigma_{nt}/v_{rms}$ $\sim$ $\sqrt{2}$, provided the wave amplitudes are unresolved in time. In other words, this relation is only valid if the period (frequency) of the wave is much smaller (higher) than the exposure time of the spectrograph. This relation is also valid if an observer performs smoothing of spectra in time. Next, we study a generalised scenario when structures oscillating in different directions and different phases happen to lie along the LOS of the observer. Due to the optically thin nature of the solar corona, the spectrum recorded by an observer will be the superposition of the spectra of all oscillating structures along the LOS. In such a scenario, $\sigma_{nt}/v_{rms}$ $\sim$ 1 irrespective of uniform or random distribution of oscillating structures along the LOS. These results are valid for circularly polarised oscillations and oscillations driven by the multi-frequency driver. We also test the scenario when wave amplitude is of the order of the $\sigma$. Taking wave amplitudes of 11$\sqrt{2}$~km~$s^{-1}$ and $\sigma$=19/$\sqrt{2}$~km~$s^{-1}$, we note about 10\% change from the limiting case in the estimated magnitude of the nonthermal line widths (see Table~\ref{table1}). To substantiate the mathematical model, we performed 3D MHD simulations and forward modeling of gravitationally stratified plasma and launched transverse MHD waves by driving the bottom boundary transversely. We note from Table~\ref{table1} that the results of the numerical simulations are in good agreement with those obtained analytically.\\
This allows us to believe that depending on the scenario, either $\sigma_{nt}$ $> \sqrt{2}$ $v_{rms}$ or $\sigma_{nt}>v_{rms}$, but never $\sigma_{nt}$= $v_{rms}/\sqrt{2}$ as used in previous studies. This raises questions about the estimation of the energy flux carried by these waves and claims made in earlier studies that the transverse wave carry enough energy to heat the solar corona using the observed nonthermal line widths.
Table~\ref{energy_table} quotes the earlier and revised (expressions derived in this study) estimates of the energy flux carried by the Alfv\'enic waves. As discussed above, the nonthermal line widths might be larger than wave amplitudes ($\sigma_{nt}>v_{rms}$) if the wave amplitude is of the order of the thermal line widths. Therefore, the wave energy flux, $F < \rho \sigma_{nt}^2 v_{A}$. Furthermore, this leads us to believe that the hotter emission lines (with large thermal line widths) are more suitable for studying large and small amplitude transverse waves.\\
We find that the non-WKB propagation of the transverse wave in a gravitationally stratified plasma can cause the rms wave amplitudes to increase more slowly with height than expected from the WKB approximation. We also report that for a million degree hot corona where scale heights are of the order of 50 Mm, low frequency transverse waves of period $\sim$ 400 s do not propagate like those predicted by the WKB theory. In such a scenario, a slow increase of wave amplitudes (hence nonthermal line widths) might be due to non-WKB nature of propagation. It should be noted that we are not ruling out wave damping in the solar atmosphere as a mechanism for the deviation of the nonthermal line width from the WKB theory. In this study we show that at least a part of leveling-off of rms wave amplitudes(hence line widths) is due to the non-WKB nature of propagation of transverse wave in the solar atmosphere.

\begin{table}[htb]
\centering
\caption{Earlier and revised estimates of the Alfv\'enic wave energy flux}  
\begin{tabular}{|c||c|c|}
\hline
Reference&Energy Flux (erg cm$^{-2} $s$^{-1}$) & Revised Energy Flux (erg cm$^{-2}$ s$^{-1}$) \\
\hline
{\citet{1990ApJ...348L..77H}}&$4.3\times10^{5}$ & $<2.15\times10^{5}$\\
{\citet{1998SoPh..181...91D}}&$3.1\times10^{5}$&$<1.52\times10^{5}$\\
{\citet{b98}}&$4.9\times10^{5}$ & $<2.45\times10^{5}$ \\
{\citet{b2009}}&$1.85\times10^{6}$&$<9.25\times10^5$\\
{\citet{2012ApJ...753...36H}}& $5.4 \times 10^{5}$&$<2.7 \times 10^{5}$\\
\hline
\end{tabular}
\label{energy_table}
\end{table}

\appendix
\section{Estimating nonthermal line widths of a period averaged spectrum}
\label{appa}
 Since second term in equation~\ref{eq2} can be neglected, we use the first term for obtaining an analytical expression. In section~\ref{sec2.1}, we mention that integrating over one period is a good assumption. Thus, we integrate over one period only. Equation~\ref{eq2} can be reformulated as follows.
\begin{equation}
\big \langle G(\mathfrak{v},t') \big \rangle _{t}=\int_{0}^{P} \frac{1}{P \sigma \sqrt{2 \pi}} \exp(-\frac{(\mathfrak{v}-v_{0}~\cos(\omega t'))^{2}}{2\sigma^{2}}) dt.
\label{eqq1}
\end{equation}
Equation~\ref{eqq1} can be written as,
\begin{equation}
\big \langle G(\mathfrak{v},t') \big \rangle _{t}=\int_{0}^{P} \frac{1}{P \sigma \sqrt{2 \pi}} Q dt.
\label{eqq2}
\end{equation}
Where $Q$ can be expanded as,
\begin{equation}
Q= 1- \frac{(\mathfrak{v} - v_{0}~\cos(\omega t'))^{2}}{2\sigma^{2}} + \frac{1}{2} \frac{(\mathfrak{v} - v_{0}~\cos(\omega t'))^{4}}{4\sigma^{4}} - \frac{1}{6} \frac{(\mathfrak{v} - v_{0}~\cos(\omega t'))^{6}}{8\sigma^{6}}  + ....
\label{eqq3}
\end{equation}
For simplicity, we consider first four terms only for the further analysis. The second term (say $A$) in Equation~\ref{eqq3} can be expanded as follows,
\begin{equation}
A=\frac{\mathfrak{v}^{2}}{2\sigma^{2}}-\frac{2\mathfrak{v}v_{0}\cos(\omega t')}{2\sigma^{2}}+\frac{v_{0}^{2} \cos^{2} (\omega t') }{2\sigma^{2}}.
\label{eqq4}
\end{equation}
Since, we integrate over one period, the integration over time of the odd powers of the cosine function vanishes. Now onwards, we will consider only even powers of the cosine function.\\
Similarly, third (say $B$) and fourth terms (say $C$) in Equation~\ref{eqq3} can be expanded to the following equations (after ignoring odd powers of the cosine function).
\begin{equation}
B=\frac{\mathfrak{v}^{4}}{8\sigma^{4}}+\frac{6\mathfrak{v}^{2} v_{0}^{2} \cos^{2} (\omega t') }{8\sigma^{4}}+\frac{v_{0}^{4} \cos^{4} (\omega t') }{8\sigma^{4}}
\label{eqq5}
\end{equation}
\begin{equation}
C=\frac{\mathfrak{v}^{6}}{48\sigma^{6}}+\frac{15\mathfrak{v}^{4} v_{0}^{2} \cos^{2} (\omega t') }{48\sigma^{6}}+\frac{15\mathfrak{v}^{2} v_{0}^{4} \cos^{4} (\omega t') }{48\sigma^{6}}+\frac{v_{0}^{6} \cos^{6} (\omega t') }{48\sigma^{6}}
\label{eqq6}
\end{equation}
The period averages of $\cos^2(\omega t')$, $\cos^{4}(\omega t')$, and $\cos^6(\omega t')$ are $\frac{1}{2}$, $\frac{3}{8}$, and $\frac{5}{16}$ respectively. Note, period average of odd powers of the cosine function vanishes. We replace cosine functions with these values and drop the integration. The Equation~\ref{eqq2} can be rearranged to the following equation after collecting the terms of same order in $\mathfrak{v}$.
\begin{equation}
\big \langle G(\mathfrak{v},t') \big \rangle _{t}=\frac{1}{\sigma \sqrt{2 \pi}}\Big[(1-\frac{v_{0}^{2}}{4\sigma^{2}}+\frac{3v_{0}^{4}}{64\sigma^{4}}-\frac{5v_{0}^{6} }{768\sigma^{6}}+...) - \frac{\mathfrak{v}^{2}}{2\sigma^{2}}(1-\frac{3 v_{0}^{2}}{4\sigma^{2}}+ \frac{15 v_{0}^{4}}{64\sigma^{4}}+ ... ) +  \frac{\mathfrak{v}^{4}}{8\sigma^{4}}(1-  \frac{15 v_{0}^{2}}{12\sigma^{2}}+...) + ... \Big]
\label{eqq7}
\end{equation}
Assuming $v_{0}^2 \ll \sigma^{2} $, Equation~\ref{eqq7} can be reformulated as follows
\begin{equation}
\big \langle G(\mathfrak{v},t') \big \rangle _{t} \approx \frac{1}{\sigma \sqrt{2 \pi}}\Big[  \exp(- \frac{v_{0}^{2}}{4\sigma^{2}}) -  \frac{\mathfrak{v}^{2}}{2\sigma^{2}} \exp(-\frac{3v_{0}^{2}}{4\sigma^{2}}) + \frac{\mathfrak{v}^{4}}{8\sigma^{4}} \exp(-\frac{15 v_{0}^{2}}{12 \sigma^{2}})+...  \Big]
\label{eqq8}
\end{equation}
\begin{equation}
\big \langle G(\mathfrak{v},t') \big \rangle _{t} \approx \frac{\exp(- \frac{v_{0}^{2}}{4\sigma^{2}})}{\sigma \sqrt{2 \pi}}\Big[ 1 -  \frac{\mathfrak{v}^{2}}{2\sigma^{2}} \exp(-\frac{v_{0}^{2}}{2\sigma^{2}}) + \frac{\mathfrak{v}^{4}}{8\sigma^{4}} \exp(-\frac{v_{0}^{2}}{\sigma^{2}})+...  \Big]
\label{eqq9}
\end{equation}
Assuming $\sigma '$=$\sigma \exp( \frac{v_{0}^{2}}{4\sigma^{2}})$, Equation~\ref{eqq9} can be written as follows,
\begin{equation}
\big \langle G(\mathfrak{v},t') \big \rangle _{t} \approx \frac{1}{\sigma ' \sqrt{2 \pi}}\Big[ 1 -  \frac{\mathfrak{v}^{2}}{2\sigma '^{2}} + \frac{\mathfrak{v}^{4}}{8\sigma '^{4}} +...  \Big].
\label{eqq10}
\end{equation}
\begin{equation}
\big \langle G(\mathfrak{v},t') \big \rangle _{t} \approx \frac{1}{\sigma ' \sqrt{2 \pi}}\exp(-\frac{\mathfrak{v}^2}{2\sigma '^2}).
\label{eqq11}
\end{equation}
We note that the period averaged spectrum under the assumption described above is a Gaussian described by Equation~\ref{eqq11}. 
The nonthermal width, $\sigma_{nt}$ of $\big \langle G(\mathfrak{v},t') \big \rangle _{t}$ can be estimated using the following relation.
\begin{equation}
\sigma_{nt}^{2} = \sigma'^2_{1/e} - \sigma^2_{1/e}=2\sigma '^2 - 2\sigma^2 = 2\sigma^2 (\exp(\frac{v_{0}^2}{2\sigma ^2} )-1)\approx v_{0}^2\approx 2v_{rms}^2
\label{eqq12}
\end{equation}
Thus, we find that the nonthermal line widths is approximately equal to the wave amplitude when $v_{0}\ll\sigma$. \\

\section{Estimating nonthermal line widths of a period and LOS averaged spectrum}
\label{appb}

In order to obtain an analytical relation between rms wave velocity and the nonthermal line widths for a period and LOS averaged spectrum assuming $v_{0} \ll \sigma$, we use Equation~\ref{eqq11} that describes the shape of a period averaged spectrum for oscillations aligned along the LOS (meaning $\theta$ is zero. See Figure~\ref{fig1} (a)). For a given $\theta$, the period averaged spectrum is described by Equation~\ref{eqq11} but replacing $\sigma '$ with $\sigma \exp(\frac{v_{0}^{2} \cos^{2}\theta}{4\sigma^{2}})$. Next, we integrate the resulting equation in $\theta$ as shown below.
\begin{equation}
\big \langle G(\mathfrak{v}) \big \rangle _{t,\theta} \approx \int_{0}^{2\pi} \frac{1}{ \sigma \exp( \frac{v_{0}^{2} \cos^2\theta'}{4\sigma^{2}} )\sqrt{2 \pi}}\exp\Big(-\frac{\mathfrak{v}^2}{2\sigma^2 \exp (\frac{v_{0}^{2} \cos^2\theta'}{2\sigma^{2}})}\Big) d\theta
\label{eqq13}
\end{equation}

\begin{equation}
\big \langle G(\mathfrak{v}) \big \rangle _{t,\theta} \approx \int_{0}^{2\pi} \frac{ \exp( -\frac{v_{0}^{2} \cos^2\theta'}{4\sigma^{2}} )}{ \sigma\sqrt{2 \pi}}\exp\Big(-\frac{\mathfrak{v}^2 \exp (-\frac{v_{0}^{2} \cos^2\theta'}{2\sigma^{2}})}{2\sigma^2}\Big) d\theta
\label{eqq13b}
\end{equation}

\begin{equation}
\big \langle G(\mathfrak{v}) \big \rangle _{t,\theta} \approx \int_{0}^{2\pi} \frac{ \exp( -\frac{v_{0}^{2} \cos^2\theta'}{4\sigma^{2}} )}{ \sigma\sqrt{2 \pi}}\Big[1-\frac{\mathfrak{v}^2 \exp (-\frac{v_{0}^{2} \cos^2\theta'}{2\sigma^{2}})}{2\sigma^2} + \frac{\mathfrak{v}^4 \exp (-\frac{v_{0}^{2} \cos^2\theta'}{\sigma^{2}})}{8\sigma^4}- ...\Big] d\theta
\label{eqq13c}
\end{equation}

\begin{equation}
\big \langle G(\mathfrak{v}) \big \rangle _{t,\theta} \approx \int_{0}^{2\pi}\Big[ \frac{ \exp( -\frac{v_{0}^{2} \cos^2\theta'}{4\sigma^{2}} )}{ \sigma\sqrt{2 \pi}} - \frac{ \exp( -\frac{v_{0}^{2} \cos^2\theta'}{4\sigma^{2}} )}{ \sigma\sqrt{2 \pi}} \frac{\mathfrak{v}^2 \exp (-\frac{v_{0}^{2} \cos^2\theta'}{2\sigma^{2}})}{2\sigma^2} + \frac{ \exp( -\frac{v_{0}^{2} \cos^2\theta'}{4\sigma^{2}} )}{ \sigma\sqrt{2 \pi}}  \frac{\mathfrak{v}^4 \exp (-\frac{v_{0}^{2} \cos^2\theta'}{\sigma^{2}})}{8\sigma^4}- ...\Big] d\theta
\label{eqq13c}
\end{equation}
Next, we expand the first term in Equation~\ref{eqq13c}, say A, and ignore higher order terms to obtain the following equation. 
\begin{equation}
A= \frac{1 -\frac{v_{0}^{2} \cos^2\theta'}{4\sigma^{2}} + ....}{ \sigma\sqrt{2 \pi}}.
\label{eqq13d}
\end{equation}
Similarly, second (say B) and third (say C) term in Equation~\ref{eqq13c} can be expanded as follows.
\begin{equation}
B=-\frac{\mathfrak{v}^2}{2\sigma^2} \frac{1 -\frac{3v_{0}^{2} \cos^2\theta'}{4\sigma^{2}} + ....}{ \sigma\sqrt{2 \pi}}.
\label{eqq13e}
\end{equation}
\begin{equation}
C=\frac{\mathfrak{v}^4}{8\sigma^4} \frac{1 -\frac{5v_{0}^{2} \cos^2\theta'}{4\sigma^{2}} + ....}{ \sigma\sqrt{2 \pi}}.
\label{eqq13f}
\end{equation}
Integrating Equations~\ref{eqq13d},~\ref{eqq13e}, and~\ref{eqq13f} in $\theta$ (replacing $\cos^2\theta'$ with $\frac{1}{2}$) and inserting these expressions back in Equation~\ref{eqq13c}, we get
\begin{equation}
\big \langle G(\mathfrak{v}) \big \rangle _{t,\theta}=\frac{1}{\sigma\sqrt{2\pi}}\Big[(1-   \frac{v_{0}^{2}}{8\sigma^{2}} + ...) -\frac{\mathfrak{v}^2}{2\sigma^2} (1 -\frac{3v_{0}^{2}}{8\sigma^{2}} + ....) +  \frac{\mathfrak{v}^4}{8\sigma^4} (1 -\frac{5v_{0}^{2}}{8\sigma^{2}} +...) - ...  \Big].
\label{eqq13g}
\end{equation}
If $v_{0}\ll\sigma$, we get
\begin{equation}
\big \langle G(\mathfrak{v}) \big \rangle _{t,\theta}=\frac{\exp(-\frac{v_{0}^2}{8\sigma^2})}{\sigma\sqrt{2\pi}}\Big[(1 - \frac{\mathfrak{v}^2}{2\sigma^2} \exp(-\frac{v_{0}^{2}}{4\sigma^{2}}) +  \frac{\mathfrak{v}^4}{8\sigma^4} \exp(-\frac{v_{0}^{2}}{2\sigma^{2}}) - ...  \Big].
\label{eqq13h}
\end{equation}
Equation~\ref{eqq13h} can be reformulated as follows.
\begin{equation}
\big \langle G(\mathfrak{v}) \big \rangle _{t,\theta} \approx \frac{1}{\sigma '' \sqrt{2 \pi}}\exp(-\frac{\mathfrak{v}^2}{2\sigma ''^2}).
\label{eqq14}
\end{equation}
Where, $\sigma ''$=$\sigma \exp( \frac{v_{0}^{2}}{8\sigma^{2}})$. Thus,
\begin{equation}
\sigma_{nt}^{2} = 2\sigma ''^2 - 2\sigma^2 = 2\sigma^2 (\exp(\frac{v_{0}^2}{4\sigma ^2} )-1)\approx \frac{v_{0}^2}{2}\approx v_{rms}^2.
\label{eqq15}
\end{equation}
The difference between $\sigma_{nt}$ computed using Equation~\ref{eqq15} and that estimated using Equation~\ref{eqq12} is a factor of 1/2 that appears because of the integration of $\cos^2\theta'$ inside the exponent in Equation~\ref{eqq13}. From this analysis, we conclude that whenever  $v_{0}\ll\sigma$, we can expand the exponential functions by keeping terms up to second order in $v_{0}$. Finally, the integration over all directions leads to a factor of 1/2 appearing in Equation~\ref{eqq15}. This is somewhat similar to the rms type averaging of spectra of different nonthermal line widths that leads to overall reduction in the $\sigma_{nt}$ by $\sqrt{2}$ in this scenario compared to that discussed in the appendix~\ref{appa}.

\acknowledgements
We thank anonymous referee for his/her valuable suggestions that has improved the manuscript. TVD and VP were supported by the European Research Council (ERC) under the European Union's Horizon 2020 research and innovation programme (grant agreement No 724326). TVD is also supported by the C1 grant TRACEspace of Internal Funds KU Leuven (number C14/19/089).




\begin{thebibliography}{}
\expandafter\ifx\csname natexlab\endcsname\relax\def\natexlab#1{#1}\fi
\providecommand{\url}[1]{\href{#1}{#1}}
\providecommand{\dodoi}[1]{doi:~\href{http://doi.org/#1}{\nolinkurl{#1}}}
\providecommand{\doeprint}[1]{\href{http://ascl.net/#1}{\nolinkurl{http://ascl.net/#1}}}
\providecommand{\doarXiv}[1]{\href{https://arxiv.org/abs/#1}{\nolinkurl{https://arxiv.org/abs/#1}}}

\bibitem[{{Arregui}(2015)}]{2015RSPTA.37340261A}
{Arregui}, I. 2015, Philosophical Transactions of the Royal Society of London
  Series A, 373, 20140261, \dodoi{10.1098/rsta.2014.0261}

\bibitem[{{Banerjee} {et~al.}(2007){Banerjee}, {Erd{\'e}lyi}, {Oliver}, \&
  {O'Shea}}]{2007SoPh..246....3B}
{Banerjee}, D., {Erd{\'e}lyi}, R., {Oliver}, R., \& {O'Shea}, E. 2007,
  \solphys, 246, 3, \dodoi{10.1007/s11207-007-9029-z}

\bibitem[{{Banerjee} {et~al.}(2009){Banerjee}, {P{\'e}rez-Su{\'a}rez}, \&
  {Doyle}}]{b2009}
{Banerjee}, D., {P{\'e}rez-Su{\'a}rez}, D., \& {Doyle}, J.~G. 2009, \aap, 501,
  L15, \dodoi{10.1051/0004-6361/200912242}

\bibitem[{{Banerjee} {et~al.}(1998){Banerjee}, {Teriaca}, {Doyle}, \&
  {Wilhelm}}]{b98}
{Banerjee}, D., {Teriaca}, L., {Doyle}, J.~G., \& {Wilhelm}, K. 1998, \aap,
  339, 208

\bibitem[{{Bemporad} \& {Abbo}(2012)}]{2012ApJ...751..110B}
{Bemporad}, A., \& {Abbo}, L. 2012, \apj, 751, 110,
  \dodoi{10.1088/0004-637X/751/2/110}

\bibitem[{{Chae} {et~al.}(1998){Chae}, {Sch{\"u}hle}, \&
  {Lemaire}}]{1998ApJ...505..957C}
{Chae}, J., {Sch{\"u}hle}, U., \& {Lemaire}, P. 1998, \apj, 505, 957,
  \dodoi{10.1086/306179}

\bibitem[{{De Moortel} \& {Pascoe}(2012)}]{2012ApJ...746...31D}
{De Moortel}, I., \& {Pascoe}, D.~J. 2012, \apj, 746, 31,
  \dodoi{10.1088/0004-637X/746/1/31}

\bibitem[{{De Pontieu} \& {McIntosh}(2010)}]{2010ApJ...722.1013D}
{De Pontieu}, B., \& {McIntosh}, S.~W. 2010, \apj, 722, 1013,
  \dodoi{10.1088/0004-637X/722/2/1013}

\bibitem[{{Del Zanna} {et~al.}(2019){Del Zanna}, {Gupta}, \&
  {Mason}}]{2019A&A...631A.163D}
{Del Zanna}, G., {Gupta}, G.~R., \& {Mason}, H.~E. 2019, \aap, 631, A163,
  \dodoi{10.1051/0004-6361/201834625}

\bibitem[{{Dolla} \& {Solomon}(2008)}]{2008A&A...483..271D}
{Dolla}, L., \& {Solomon}, J. 2008, \aap, 483, 271,
  \dodoi{10.1051/0004-6361:20077903}

\bibitem[{{Doschek} {et~al.}(1976{\natexlab{a}}){Doschek}, {Feldman}, \&
  {Bohlin}}]{1976ApJ...205L.177D}
{Doschek}, G.~A., {Feldman}, U., \& {Bohlin}, J.~D. 1976{\natexlab{a}}, \apjl,
  205, L177, \dodoi{10.1086/182118}

\bibitem[{{Doschek} {et~al.}(1976{\natexlab{b}}){Doschek}, {Feldman},
  {Vanhoosier}, \& {Bartoe}}]{1976ApJS...31..417D}
{Doschek}, G.~A., {Feldman}, U., {Vanhoosier}, M.~E., \& {Bartoe}, J.-D.~F.
  1976{\natexlab{b}}, \apjs, 31, 417, \dodoi{10.1086/190386}

\bibitem[{{Doyle} {et~al.}(1998){Doyle}, {Banerjee}, \&
  {Perez}}]{1998SoPh..181...91D}
{Doyle}, J.~G., {Banerjee}, D., \& {Perez}, M.~E. 1998, \solphys, 181, 91,
  \dodoi{10.1023/A:1005019931323}

\bibitem[{{Feldman} {et~al.}(1976){Feldman}, {Doschek}, {Vanhoosier}, \&
  {Purcell}}]{1976ApJS...31..445F}
{Feldman}, U., {Doschek}, G.~A., {Vanhoosier}, M.~E., \& {Purcell}, J.~D. 1976,
  \apjs, 31, 445, \dodoi{10.1086/190387}

\bibitem[{{Ferraro}(1955)}]{1955RSPSA.233..310F}
{Ferraro}, V.~C.~A. 1955, Proceedings of the Royal Society of London Series A,
  233, 310, \dodoi{10.1098/rspa.1955.0267}

\bibitem[{{Goldstein}(1978)}]{1978ApJ...219..700G}
{Goldstein}, M.~L. 1978, \apj, 219, 700, \dodoi{10.1086/155829}

\bibitem[{{Gupta} {et~al.}(2019){Gupta}, {Del Zanna}, \&
  {Mason}}]{2019A&A...627A..62G}
{Gupta}, G.~R., {Del Zanna}, G., \& {Mason}, H.~E. 2019, \aap, 627, A62,
  \dodoi{10.1051/0004-6361/201935357}

\bibitem[{{Hahn} {et~al.}(2012){Hahn}, {Landi}, \&
  {Savin}}]{2012ApJ...753...36H}
{Hahn}, M., {Landi}, E., \& {Savin}, D.~W. 2012, \apj, 753, 36,
  \dodoi{10.1088/0004-637X/753/1/36}

\bibitem[{{Hahn} \& {Savin}(2014)}]{2014ApJ...795..111H}
{Hahn}, M., \& {Savin}, D.~W. 2014, \apj, 795, 111,
  \dodoi{10.1088/0004-637X/795/2/111}

\bibitem[{{Hassler} {et~al.}(1990){Hassler}, {Rottman}, {Shoub}, \&
  {Holzer}}]{1990ApJ...348L..77H}
{Hassler}, D.~M., {Rottman}, G.~J., {Shoub}, E.~C., \& {Holzer}, T.~E. 1990,
  \apjl, 348, L77, \dodoi{10.1086/185635}

\bibitem[{{Heinemann} \& {Olbert}(1980)}]{1980JGR....85.1311H}
{Heinemann}, M., \& {Olbert}, S. 1980, \jgr, 85, 1311,
  \dodoi{10.1029/JA085iA03p01311}

\bibitem[{{Hollweg}(1972)}]{1972ApJ...177..255H}
{Hollweg}, J.~V. 1972, \apj, 177, 255, \dodoi{10.1086/151703}

\bibitem[{{Hollweg}(1973)}]{1973ApJ...181..547H}
---. 1973, \apj, 181, 547, \dodoi{10.1086/152072}

\bibitem[{{Hollweg}(1978)}]{1978SoPh...56..305H}
---. 1978, \solphys, 56, 305, \dodoi{10.1007/BF00152474}

\bibitem[{{Hollweg}(1981)}]{1981SoPh...70...25H}
---. 1981, \solphys, 70, 25, \dodoi{10.1007/BF00154391}

\bibitem[{{Klimchuk}(2006)}]{2006SoPh..234...41K}
{Klimchuk}, J.~A. 2006, \solphys, 234, 41, \dodoi{10.1007/s11207-006-0055-z}

\bibitem[{{Magyar} {et~al.}(2017){Magyar}, {Van Doorsselaere}, \&
  {Goossens}}]{2017NatSR...714820M}
{Magyar}, N., {Van Doorsselaere}, T., \& {Goossens}, M. 2017, Scientific
  Reports, 7, 14820, \dodoi{10.1038/s41598-017-13660-1}

\bibitem[{{Magyar} {et~al.}(2019){Magyar}, {Van Doorsselaere}, \&
  {Goossens}}]{2019ApJ...882...50M}
---. 2019, \apj, 882, 50, \dodoi{10.3847/1538-4357/ab357c}

\bibitem[{{McIntosh} \& {De Pontieu}(2012)}]{2012ApJ...761..138M}
{McIntosh}, S.~W., \& {De Pontieu}, B. 2012, \apj, 761, 138,
  \dodoi{10.1088/0004-637X/761/2/138}

\bibitem[{{McIntosh} {et~al.}(2011){McIntosh}, {Leamon}, \& {De
  Pontieu}}]{2011ApJ...727....7M}
{McIntosh}, S.~W., {Leamon}, R.~J., \& {De Pontieu}, B. 2011, \apj, 727, 7,
  \dodoi{10.1088/0004-637X/727/1/7}

\bibitem[{{Morton} {et~al.}(2015){Morton}, {Tomczyk}, \&
  {Pinto}}]{2015NatCo...6E7813M}
{Morton}, R.~J., {Tomczyk}, S., \& {Pinto}, R. 2015, Nature Communications, 6,
  7813, \dodoi{10.1038/ncomms8813}

\bibitem[{{O'Shea} {et~al.}(2005){O'Shea}, {Banerjee}, \&
  {Doyle}}]{2005A&A...436L..35O}
{O'Shea}, E., {Banerjee}, D., \& {Doyle}, J.~G. 2005, \aap, 436, L35,
  \dodoi{10.1051/0004-6361:200500120}

\bibitem[{{Pant} {et~al.}(2019){Pant}, {Magyar}, {Van Doorsselaere}, \&
  {Morton}}]{2019ApJ...881...95P}
{Pant}, V., {Magyar}, N., {Van Doorsselaere}, T., \& {Morton}, R.~J. 2019,
  \apj, 881, 95, \dodoi{10.3847/1538-4357/ab2da3}

\bibitem[{{Parnell} \& {De Moortel}(2012)}]{2012RSPTA.370.3217P}
{Parnell}, C.~E., \& {De Moortel}, I. 2012, Philosophical Transactions of the
  Royal Society of London Series A, 370, 3217, \dodoi{10.1098/rsta.2012.0113}

\bibitem[{{Pascoe} {et~al.}(2019){Pascoe}, {Smyrli}, \& {Van
  Doorsselaere}}]{2019ApJ...884...43P}
{Pascoe}, D.~J., {Smyrli}, A., \& {Van Doorsselaere}, T. 2019, \apj, 884, 43,
  \dodoi{10.3847/1538-4357/ab3e39}

\bibitem[{{Porth} {et~al.}(2014){Porth}, {Xia}, {Hendrix}, {Moschou}, \&
  {Keppens}}]{2014ApJS..214....4P}
{Porth}, O., {Xia}, C., {Hendrix}, T., {Moschou}, S.~P., \& {Keppens}, R. 2014,
  \apjs, 214, 4, \dodoi{10.1088/0067-0049/214/1/4}

\bibitem[{{Tian} {et~al.}(2011{\natexlab{a}}){Tian}, {McIntosh}, \& {De
  Pontieu}}]{2011ApJ...727L..37T}
{Tian}, H., {McIntosh}, S.~W., \& {De Pontieu}, B. 2011{\natexlab{a}}, \apjl,
  727, L37, \dodoi{10.1088/2041-8205/727/2/L37}

\bibitem[{{Tian} {et~al.}(2011{\natexlab{b}}){Tian}, {McIntosh}, {Habbal}, \&
  {He}}]{2011ApJ...736..130T}
{Tian}, H., {McIntosh}, S.~W., {Habbal}, S.~R., \& {He}, J. 2011{\natexlab{b}},
  \apj, 736, 130, \dodoi{10.1088/0004-637X/736/2/130}

\bibitem[{{Tian} {et~al.}(2012){Tian}, {McIntosh}, {Wang}, {Ofman}, {De
  Pontieu}, {Innes}, \& {Peter}}]{2012ApJ...759..144T}
{Tian}, H., {McIntosh}, S.~W., {Wang}, T., {et~al.} 2012, \apj, 759, 144,
  \dodoi{10.1088/0004-637X/759/2/144}

\bibitem[{{Tomczyk} \& {McIntosh}(2009)}]{2009ApJ...697.1384T}
{Tomczyk}, S., \& {McIntosh}, S.~W. 2009, \apj, 697, 1384,
  \dodoi{10.1088/0004-637X/697/2/1384}

\bibitem[{{Tu} {et~al.}(1998){Tu}, {Marsch}, {Wilhelm}, \&
  {Curdt}}]{1998ApJ...503..475T}
{Tu}, C.-Y., {Marsch}, E., {Wilhelm}, K., \& {Curdt}, W. 1998, \apj, 503, 475,
  \dodoi{10.1086/305982}

\bibitem[{{van Ballegooijen} {et~al.}(2017){van Ballegooijen}, {Asgari-Targhi},
  \& {Voss}}]{2017ApJ...849...46V}
{van Ballegooijen}, A.~A., {Asgari-Targhi}, M., \& {Voss}, A. 2017, \apj, 849,
  46, \dodoi{10.3847/1538-4357/aa9118}

\bibitem[{Van~Doorsselaere {et~al.}(2016)Van~Doorsselaere, Antolin, Yuan,
  Reznikova, \& Magyar}]{10.3389/fspas.2016.00004}
Van~Doorsselaere, T., Antolin, P., Yuan, D., Reznikova, V., \& Magyar, N. 2016,
  Frontiers in Astronomy and Space Sciences, 3, 4,
  \dodoi{10.3389/fspas.2016.00004}

\bibitem[{{Walker}(2005)}]{walker}
{Walker}, A. 2005, {Magnetohydrodynamic Waves in Geospace: The Theory of Ulf
  Waves and Their Interaction with Energetic Particles in the Solar-Terrestrial
  Environment. Series in Plasma Physics (London: Institute of Physics)}

\bibitem[{{Walsh} \& {Ireland}(2003)}]{2003A&ARv..12....1W}
{Walsh}, R.~W., \& {Ireland}, J. 2003, \aapr, 12, 1,
  \dodoi{10.1007/s00159-003-0021-9}

\bibitem[{{Weberg} {et~al.}(2020){Weberg}, {Morton}, \&
  {McLaughlin}}]{2020ApJ...894...79W}
{Weberg}, M.~J., {Morton}, R.~J., \& {McLaughlin}, J.~A. 2020, \apj, 894, 79,
  \dodoi{10.3847/1538-4357/ab7c59}

\end{thebibliography}
\end{document}